# An EEG-based approach for Parkinson's disease diagnosis using capsule network

Shujie Wang[1,a], Gongshu Wang[1], Guangying Pei[1,b,*] and Tianyi Yan[1]

[1]School of Life Science, Beijing Institute of Technology, Haidian District, Beijing, China
[a]wangshujie@bit.edu.cn
[b]pei_guangying@bit.edu.cn

*Abstract*—As the second most common neurodegenerative disease, Parkinson's disease has caused serious problems worldwide. However, the pathology and mechanism of PD are still unclear, and a systematic early diagnosis and treatment method for PD has not yet been established. Many patients with PD have not been diagnosed or misdiagnosed. In this paper, we proposed an EEG-based approach to diagnosing Parkinson's disease. The frequency band energy of the electroencephalogram (EEG) signal was mapped to the 2-dimensional image by using the interpolation method, and identified classification based on the capsule network (CapsNet) and achieved 89.34% classification accuracy for short-term EEG sections. By comparing the individual classification accuracy of different EEG frequency bands, we found that the gamma band has the highest accuracy, providing potential feature targets for the early diagnosis and clinical treatment of PD.

*Keywords-Parkinson's disease; machine learning; deep learning; electroencephalograph; capsule network*

## I Introduction

Parkinson's disease (PD) is the second most common neurodegenerative disease in the world. It is common in the elderly and affects 2–3% of people over 60 years of age [1,2]. According to the Global Burden of Disease study, neurological diseases are currently the major source of disability globally. a Considering the age-standardized prevalence, disability, and mortality. PD is the fastest growing disease of these diseases. From 1990 to 2016, the number of people with Parkinson's disease worldwide increased by 145% to 6.06 million [3].

The main clinical manifestations of PD include slow movement, limb stiffness, static tremor, and gait disorder. Many patients have sleep disorders, and anxiety is present. To date, there are no effective treatment means, mainly relying on drug intervention and comprehensive treatment to delay the progress of the disease, but the efficacy of drugs in the middle and late stages gradually declines, and complications gradually appear. Although surgical treatment can be used as a means of middle and late PD treatment, limited clinical application is applied due to invasive treatment and high cost [4]. Therefore, the early diagnosis of PD is particularly important.

However, the pathology and mechanism of PD are still unclear, and no systematic early diagnosis and treatment of PD have been established. Many patients with PD are not diagnosed or misdiagnosed, resulting in a large number of actual PD patients being unable to receive the corresponding treatment [5].

The difficulty in early diagnosis of PD is the inconspicuous characteristics and susceptibility to confusion with other diseases [6]. Therefore, how to effectively extract the features in the input signal and improve the accuracy of Parkinson's disease is an important difficulty in the diagnosis of PD.

Electroencephalogram (EEG) can record the spontaneous, rhythmic electrical activity of brain cells with good temporal resolution, is relatively easy to collect, and is widely used in the detection of neurological diseases. Studies have shown that the most important change in the EEG of PD patients is the change of main wave frequency (dominant frequency, DF) from the alpha band to the slower high theta band, that is, EEG slows down [7]. EEG can be used for the detection of non-motor symptoms, such as an EEG study in PD patients with depressive symptoms, showing increased activity of the delta and the theta band [8]. Continuous motion task completion is associated with specific EEG bands, where motion preparation corresponds to the beta band, motion control and execution correspond to the gamma band, and processing of the conflict signal corresponds to the theta band [9,10]. As the main rhythm in the cortical-spinal system, beta band is important in coordinating motor function, and motor abnormalities in PD may cause increased high beta band coherence in the sensorimotor cortical-subcortical region [11]. Combined, past studies have found that EEG can more objectively and quantitatively represent the disease process and symptoms of PD patients, which can effectively guide the early diagnosis of PD [12-14]. Among these models, convolutional neural networks (CNN) have shown superiority in physiological signal recognition [15]. However, the pooling layer in CNN ignores much information because of its data compression functions and static routing, and a capsule network (CapsNet) can solve this problem [16]. CapsNet considers the spatial relationships between features, simulates the human brain learning process, and achieves the best results on the Modified National Institute of Standards and Technology database (MNIST database).

In this paper, we propose an EEG-based approach for PD diagnosis with a capsule network. First, we use the interpolation method to project the energy features of different frequency bands of the EEG to the corresponding spatial positions to construct the EEG map, in which the raw signal of the EEG is converted into a two-dimensional (2D) image. Then input the EEG map into CapsNet for training and testing. To the best of our knowledge, this is the first attempt to diagnose Parkinson's disease with a capsule network.

TABLE I. DEMOGRAPHIC DATA.

|  | PD | HC | P-value |
|---|---|---|---|
| N (sex ratio M/F) | 55 (29/26) | 30 (15/15) | 0.813 |
| Age (SD), year | 59.82 (7.33) | 57.73 (7.63) | 0.226 |
| MMSE (SD) | 27.35 (1.71) | 27.53 (2.50) | 0.888 |
| Education (SD), year | 4.24 (0.94) | 3.90 (1.22) | 0.198 |
| MDS-UPDRS III (SD) | 30.42 (12.52) | - | - |
| MoCA (SD) | 25.52 (3.27) | - | - |
| H&Y (SD) | 2.25 (0.49) | - | - |

MMSE, Mini-Mental State Exam; MDS-UPDRS III, Movement Disorders Society-Unified Parkinson's Disease Rating Scale-Part III; MoCA, Beijing version of the Montreal Cognitive Assessment; H&Y, Hoehn & Yahr stage;

## II MATERIALS AND METHODS

### A. Participants

Eighty-five participants were recruited from November 2019 to January 2021 in the present study. Fifty-five nondemented Parkinson's disease patients were recruited from the Neurological Rehabilitation Center of Beijing Rehabilitation Hospital Affiliated to Capital Medical University, and 30 healthy controls (HC) were recruited by the local community recruiting ads. Demographic and clinical details are summarized in Table 1. This study was approved by the Ethics Committee of the Beijing Rehabilitation Hospital Affiliated with Capital Medical University and Aerospace Central Hospital following the Declaration of Helsinki, and all participants provided informed written consent before the experiment.

### B. EEG data collection and processing

Using the BP Company 32 Guide EEG Acquisition Equipment (Brain Products, Germany) for EEG data acquisition, EEG acquisition was based on the International 10/20 system. EEG was collected from 30 channels (FP1, FP2, F7, F3, Fz, F4, F8, T3, C3, Cz, C4, T4, T5, P3, Pz, P4, T, T6, FT9, FC5, FC1, CP5, CP1, Oz, CP6, CP2, FT10, FC6, FC2, O1, and O2). The reference electrode was set as the left and right papillary electrodes (TP9 and TP10). The grounding electrode was placed in front of it. The sampling rate was 1,000 Hz. The scalp impedance was reduced to 5 kΩ at the acquisition. Eye-open and closed resting EEG signals were collected simultaneously for each subject for 15 min each.

EEG signal preprocessing mainly includes the following steps: data preview, electrode positioning, filtering, independent component analysis (ICA), artifact deletion, and segmentation. First, a double electrode reference was used to reduce the error and brain hemisphere effect caused by the single reference electrode. Second, filtering the data from the high-frequency band and working frequency interference baseline drift by 0.5 Hz high-pass and 45 Hz low-frequency FIR filter. Third, ICA is used to remove noise components such as electromyogram (EMG) and electrooculogram (EOG) to obtain relatively pure EEG signals. Finally, the preprocessed EEG data were split into 5 s + 5 s epochs (open + close). To ensure that the number of epochs was the same between subjects, 90 valid epochs without artifacts were selected from each PD subject, and 165 valid epochs without artifacts were selected from each HC subject. Therefore, we obtained 4950 epochs for the PD group and 4950 epochs for the HC group, 9900 epochs in total.

Considering the inherent structure of the data in space, frequency, and time, we used a method proposed by Pouya Bashivan et al. to transform the measurements into a 2-D image to save the spatial structure and used multiple color channels to represent the spectral dimension [17]. As shown in Figure 1, for each epoch of EEG data, the energy features of theta (4-8 Hz), alpha (8-13 Hz), beta (13-30 Hz), and gamma (30-45 Hz) frequency bands in each channel were calculated by the Welch method, and the three-dimensional (3D) coordinates of electrode were extracted from the equipment information. Azimuthal equidistant projection (AEP) was used to project the electrode position to the 2D plane. Finally, the scattering power measurement on the scalp was interpolated for an EEG feature image of 8 (four-band energy characteristics and two eye-open states) ×32×32 (32×32 grid). Thus, the topographic features of the EEG spectrum energy were retained, which was more conducive to the input of the subsequent deep learning model. Finally, for k-fold cross-validation, we randomly divided each group of EEG feature images into five parts, and each part of the data contained data from 6 HC individuals and 11 PD individuals.

### C. Capsule network

A capsule network is a neural network instead of scalar neurons of a convolutional neural network, which can better capture and characterize the relationship between the characteristic attributes of the input signal (such as relative location, scale, direction, etc). The CapsNet model retains the convolutional properties of the convolutional neural network, creates higher-level capsules to cover a larger region of the image, and does not lose the exact location information within the region, while combining the capsules with dynamic routing, improving the efficiency of the capsule network.

Unlike the scalar neurons in the convolutional neural network, capsule j (vector neuron j) in the capsule network uses a matrix $W_{ij}$ to process the input vector $u_i$ into new input vectors, as shown in (1).

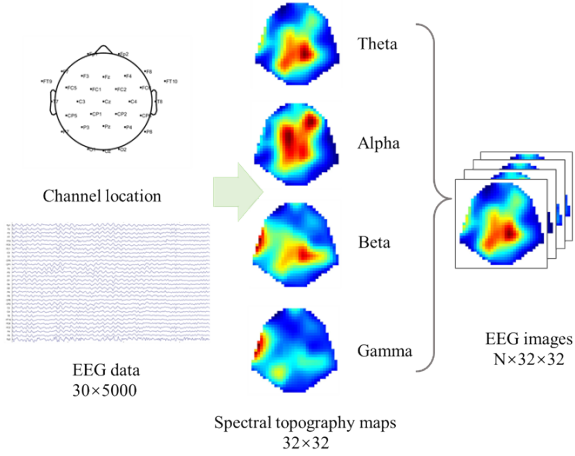

Figure 1. The method of 2D EEG image generating by EEG data

$$\hat{u}_{j|i} = W_{ij} u_i \quad (1)$$

The input vector $\hat{u}_{j|i}$ is then multiplied by the weight $c_{ij}$, and the total input of capsule j is obtained after the sum of the weighted vector $\hat{u}_{j|i}$, as shown in (2).

$$S_j = \sum_i c_{ij} \hat{u}_{j|i} \quad (2)$$

Using a nonlinear "squashing" function as activation to ensure that the capsule output vector $v_j$ is between 0-1, $v_j$ could represent the probability that the entity represented by the capsule is present in the current input, as shown in (3).

$$v_j = \frac{\|s_j\|^2}{1 + \|b_j\|^2 \|s_j\|} \frac{s_j}{} \quad (3)$$

The weight $c_{ij}$ in (2) is the coupling coefficient determined by the iterative dynamic routing process. The sum of the coupling coefficient between all k capsules of capsule i and the upper layer is 1. $c_{ij}$ is decided by the "Routing softmax", as shown in (4).

$$c_{ij} = \frac{\exp(b_{ij})}{\sum_k \exp(b_{ik})} \quad (4)$$

In formula (3), $b_{ij}$ is the log-prior probability that capsule i should be coupled to capsule j. $b_{ij}$ obtained by distinction learning, as with all other weights, depends on the location and type of the two capsules rather than the current input. For the first capsule layer, the initial value of $b_{ij}$ is 0 for capsule I in layer l and capsule j in layer (l+1), as shown in (5).

$$b_{ij} \leftarrow b_{ij} + \hat{u}_{j|i} \cdot v_j \quad (5)$$

### D. Proposed method

As shown in Figure 2. The specific implementation scheme is to obtain the EEG feature image of N (it turns on the frequency band energy features and eye-open states) ×32×32 (32×32 grid) as input and then passes through the standard convolutional layer, main capsule layer, and digital capsule layer.

The first layer is an ordinary CNN layer, which acts as pixel-level local feature detection. The input is N×32×32 in size, and the first layer uses 256 convolutional 9×9 cores with a stride of 1 and ReLU activation, resulting in an output matrix size of 24×24×256.

The second layer is the main capsule layer (Primarycaps), which can be understood as a stack of 8 parallel conventional convolutional layers with 8×32 convolution 9×9 cores with a step size of 2, yielding 8 output matrices of 8×8×1×32, changing the 3 D output tensor 6×6×1×32 under the traditional convolution to the 4 D output tensor 6×6×8×32, that is, each calculated output is a vector of length 8.

The third digital capsule layer (Digitcaps) spreads and routs updates based on the vector output of the second layer. The second layer outputs a total of 8×8×32=2048 vectors, each with a dimension of 8, namely, 2048 capsule units in layer i. While the third layer j has 2 standard capsule units, the output vector for each capsule has 16 elements. The number of capsule cells in the previous layer is 2048, so there will be 2048×2 cells, each $W_{ij}$ with dimensions of 8×16. When the predicted vectors are multiplied by their counterparts, we have 2048×2 coupling coefficients $c_{ij}$, and the corresponding weighted sum yields 2 input vectors of 16 × 1. The input vector is input into the "squashing nonlinear function to obtain the final output vector, where the length indicates the probability of identification as a category.

## III  RESULTS & DISCUSSION

In this study, we compared the CapsNet classification results with input among four kinds of frequency bands and evaluated the contribution of different frequency bands to the model. As shown in Table 2, with a single band feature, we obtained a 5-fold cross-validated accuracy of 83.85% for the gamma band,

82.48% for the beta band, 82.58% for the alpha band, and 82.49% for the theta band. Therefore, the gamma band might play an important role in Parkinson's pathology. As shown in Table 3, we obtained a significant increase in accuracy of 89.34% for all band feature inputs (four frequency band energy features and two eye-open states), which has exceeded the multi-kernel SVM algorithm (accuracy of 88.99%) proposed by our team before [14]

## IV  CONCLUSION

In this paper, we proposed a method to diagnose Parkinson's disease by EEG-based CapsNet. As a neural network model, this method has exceeded the classification accuracy of the SVM model in a small sample (less than 100) and has the potential to identify the electrophysiological signal characteristics of PD and establish a dynamic mapping model of Parkinson's disease patients. which can serve as an effective supplement to the current clinical diagnosis of PD. Moreover, the good performance of the gamma band indicates that we need to pay attention to its role in Parkinson's pathology.

ACKNOWLEDGMENT

This work was supported by the National Natural Science Foundation of China (Grant Numbers U20A20191, 61727807, 82071912, 12104049), the Beijing Municipal Science & Technology Commission (Grant Number Z201100007720009),

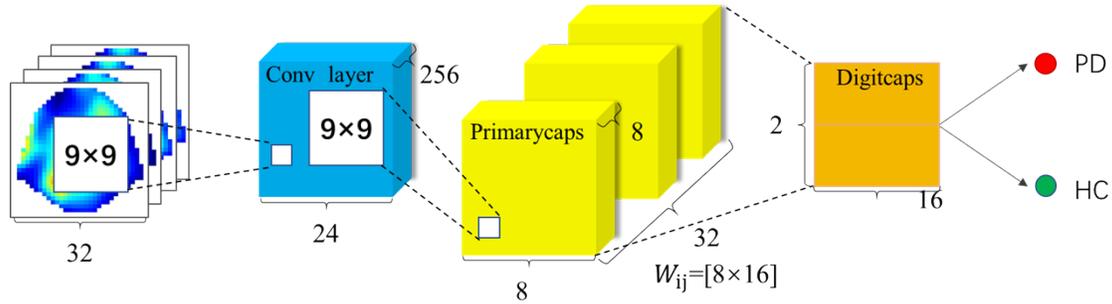

Figure 2. Overall structure of the proposed method

TABLE II. THE CLASSIFICATION PERFORMANCE OF DIFFERENT EEG FREQUENCY BANDS INPUT

| Feature | Accuracy | Sensitivity | Specificity |
|---|---|---|---|
| Theta | 82.49±3.07% | 78.71±6.07% | 86.28±6.22% |
| Alpha | 82.58±3.76% | 78.40±4.40% | 86.75±5.75% |
| Beta | 82.48±3.69% | 75.84±3.21% | 89.13±7.83% |
| Gamma | 83.85±4.76% | 82.16±5.23% | 85.54±8.05% |
| Total | 89.34±4.06% | 86.88±4.10% | 91.83±6.76% |

TABLE III. THE CLASSIFICATION PERFORMANCE OF DIFFERENT APPROACHES

| Feature | Accuracy | Sensitivity | Specificity |
|---|---|---|---|
| SVM | 88.99±4.11% | 86.45±7.14% | 91.54±3.89% |
| CapsNet | 89.34±4.06% | 86.88±4.10% | 91.83±6.76% |

the Fundamental Research Funds for the Central Universities (2021CX11011), and the China Postdoctoral Science Foundation (Grant Number 2020TQ0040).